\documentclass[%
 reprint,
superscriptaddress,
 amsmath,amssymb,
 aip
 ]{revtex4-1}

\usepackage{blindtext}
\usepackage[normalem]{ulem}
\usepackage[dvipsnames]{xcolor}
\usepackage{soul}

\usepackage{amsmath,amssymb,amsfonts}
\usepackage{algorithmic}
\usepackage{graphicx}
\usepackage{textcomp}
\usepackage{xcolor}
\usepackage{pifont}
\usepackage[hidelinks,colorlinks=true,linkcolor=blue,citecolor=blue,urlcolor=black]{hyperref}

\usepackage{cleveref}
\usepackage{flushend}
\usepackage{amssymb,amsmath,amsthm,enumitem}
\usepackage{float}

\begin{document}
\title{Retrieving Effective Acoustic Impedance and Refractive Index for Size Mismatch Samples}

\author{Mohammad Javad Khodaei}
\thanks{A.M. and M.J.K contributed equally to this work.}
\affiliation{Department of Mechanical and Industrial Engineering, Northeastern University, Boston, MA, 02115, USA.}

\author{Amin Mehrvarz}
\thanks{A.M. and M.J.K contributed equally to this work.}
\affiliation{Department of Mechanical and Industrial Engineering, Northeastern University, Boston, MA, 02115, USA.}

\author{Reza Ghaffarivardavagh}
\affiliation{MIT Media Lab, Massachusetts Institute of Technology, Cambridge, MA 02139, USA.}
\author{Nader Jalili}
\email{njalili@ua.edu}
\affiliation{Department of Mechanical Engineering, University of Alabama, Tuscaloosa, AL, 35487, USA.}
\affiliation{Department of Mechanical and Industrial Engineering, Northeastern University, Boston, MA, 02115, USA.}

\begin{abstract}
In this paper, we have presented an analytical solution to extract the effective properties of acoustic metamaterials from the measured complex transmission and reflection coefficients when the metamaterial and impedance tube have different sizes. We have first modeled this problem as a bilayer metamaterial located inside a duct and treated the air gap as a separate domain. Then we have mathematically proved that the effective properties of acoustic metamaterial can be obtained by solving a set of eight linear equations when the dimensions are known. Finally, we have evaluated the proposed method with results from numerical simulations. It is shown that the proposed method can calculate the effective refractive index and impedance with an error of below 1\%. This method provides an efficient approach to analyzing the effective properties of acoustic metamaterials of various sizes.
\end{abstract}


\maketitle 

\section{Introduction} \label{sec1}
Acoustic metamaterials are engineered structures in subwavelength scales with incredible acoustic properties. In these structures, effective properties like density, bulk modulus, impedance, and refractive index determine their functionalities in different applications. The effective acoustic properties of acoustic metamaterials are a function of their geometries and shapes rather than their materials. The most-reported acoustic metamaterials in the literature are (i) coiling-up space \cite{li2013reflected, ghaffarivardavagh2018horn}, (ii) Helmholtz-resonator-like \cite{li2015metascreen}, and (iii) membrane-type structures \cite{chen2017magnetic, tang2019voltage, liu2020magnetically}. These 2D \cite{qian2019reflected} and 3D \cite{al2016high} structures have shown strong abilities in modulating the reflected \cite{li2014experimental, guo2018manipulating, li2019reflection} and transmitted \cite{xie2014wavefront, chen2018broadband, wang2019tunable} acoustic waves to improve the wave manipulation \cite{fu2019multifunctional, fang2019acoustic, xu2019spatial, li2019modulation, ghaffarivardavagh2019tailoring, khodaei2020non}, and sound absorption and attenuation \cite{zhu2019broadband, ghaffarivardavagh2019ultra, khodaei2021acoustic}. 

\begin{figure}[!t]
    \centering
    \includegraphics[width=3.4in,scale=1]{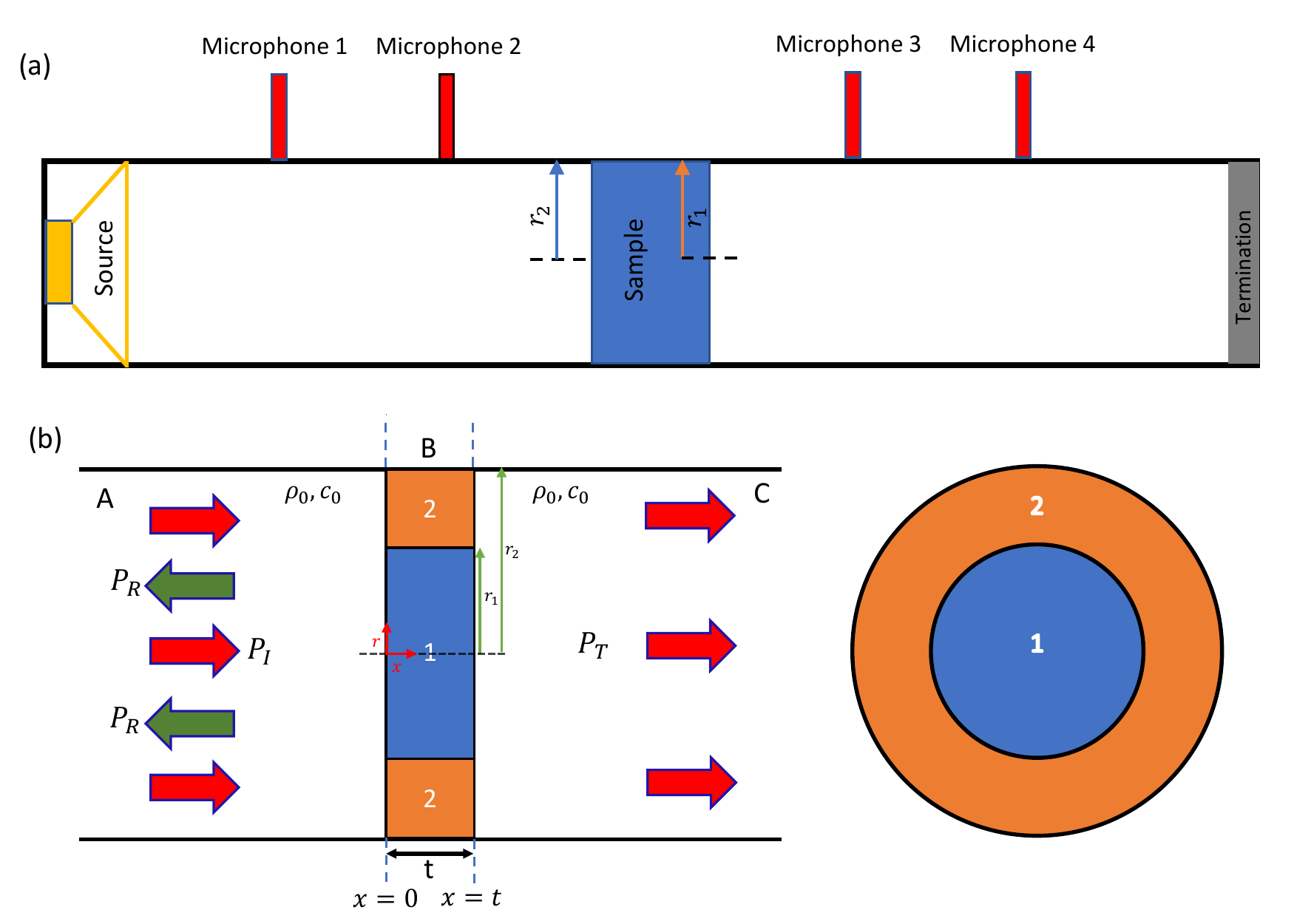}
    \caption{a) Schematic of a conventional impedance tube, b) A bilayer metamaterial located in a duct.}
    \label{fig1}
\end{figure}

Characterizing these acoustic metamaterials and extracting their effective properties are always in the field of interest of researchers. The effective properties of materials can be calculated from impedance tube measurements. An impedance tube is a test setup to measure complex reflection and transmission coefficients of materials. As shown in Fig.\ref{fig1}a, the material is located in the middle of a tube in this method. The sound source generates the acoustic incident wave, and the microphones measure the acoustic pressure in different locations. Different techniques like the transfer matrix and wave decomposition methods can be employed to calculate the materials' complex reflection and transmission coefficients from the measured acoustic pressures \cite{song2000transfer, bolton2006measurement, salissou2009general, ho2005measurements}. The effective properties of the materials can be obtained later from the calculated complex reflection and transmission coefficients utilizing a well-known retrieving method \cite{fokin2007method}. A fundamental assumption in this method is that the material fills the tube cross-section fully and, indeed, the material and tube have the same sizes (i.e., $r_1=r_2$). Impedance tube has standard sizes, and the samples of the materials must be built with these sizes to extract the effective properties utilizing the retrieving method. 

Size does not affect the effective properties of simple materials like ceramic, glass, and wood, and new samples with different sizes can be built to match the tube and sample sizes. However, the acoustic metamaterials in which their effective properties are a function of their sizes and shapes cannot be built with the standard sizes, and researchers build custom impedance tubes to fit the tube and sample sizes. This solution makes the impedance tube measurements usable in the retrieving method but requires a unique impedance tube for each acoustic metamaterial, which is not desirable.

In this paper, we propose an analytical method to extract the effective properties when the material and tube do not have the same sizes and there is a gap in the cross-section. To this end, as shown in Fig.~\ref{fig1}b, we first assume that a bilayer metamaterial (air gap and metamaterial) is located inside a duct where the radius of metamaterial ($r_1$) is less than the radius of duct ($r_2$). Subsequently, we employ the transfer matrix method (TMM) and the solution of the Helmholtz equation for a duct and mathematically prove that the effective refractive index ($n$) and impedance ($z$) can be calculated by solving eight linear equations. Eventually, we simulate the acoustic wave propagation inside the duct and bilayer the metamaterial utilizing the finite element method (FEM) and evaluate the proposed method by data from the numerical simulation. It is demonstrated that the proposed method in this paper successfully calculates the refractive index ($n$) and impedance ($z$) with an accuracy of 99\%. The presented method extends the previous method and facilitates the study of acoustic metamaterials by providing an analytical solution to extract their effective acoustic properties when the impedance tube and material do not possess the same sizes. 

\section{Proposed Method to Extract the Effective Properties} \label{sec2}
The transfer matrix method (TMM) has shown a strong ability to accurately model planar acoustic waves propagation throughout the duct and metamaterials and determine the complex transmission and reflection coefficients. In this section, the transfer matrix method (TMM) is utilized to propose an analytical retrieval method to extract the effective properties of acoustic metamaterial like refractive index ($n$) and impedance ($z$) when the metamaterial and duct are not at the same size.

As shown in Fig.~\ref{fig1}, three regions are assumed in this problem. Regions $A$ and $C$ include the incoming, outcoming, and reflected pressures ($P_I$, $P_T$, $P_R$), which are located before and after the bilayer metamaterial. In these two regions, air flows with density and velocity of $\rho_0$ and $c_0$. Region B includes the bilayer metamaterial and has two parts. Part 1 is the unknown metamaterial with an effective refractive index and effective impedance of $n_1$ and $z_1$, which are under investigation in this paper. Part 2 represents the air gap, which is the result of the discrepancy between the duct and metamaterial size, with an effective refractive index and effective impedance of $n_2=1$ and $z_2=\rho_0 c_0/\pi(r_2^2-r_1^2)$. In this paper, the hard boundary assumption is taken in the interface of parts 1 and 2, which results in no acoustic wave propagation between these two parts. Also, the thermoviscous effects can be ignored for air-borne sounds \cite{jia2018metasurface, tong2019high}.

The complex transmission ($T=P_T/P_I$) and reflection ($R=P_R/P_I$) coefficients can be written as \cite{ghaffarivardavagh2019ultra}:
\begin{equation} \label{eq1}
    T=\frac{2}{M_{11}+M_{12}/\alpha+\alpha M_{21}+M_{22} }
\end{equation}
\begin{equation}\label{eq2}
    R=\frac{M_{11}+\frac{M_{12}}{\alpha}-\alpha M_{21}-M_{22}}{M_{11}+\frac{M_{12}}{\alpha}+\alpha M_{21}+M_{22} }
\end{equation}
where $M_{ij}$ denotes the elements of the transfer matrix and will be calculated from measured $T$ and $R$ (see Supplementary Note A), and $\alpha$ is $\rho_0 c_0$. The transfer matrix of the transverse bilayer metamaterial (part B) can be written as: 
\begin{equation}\label{eq3}
    \begin{bmatrix}
        P_{x0}\\
        u_{x0}
    \end{bmatrix}=
    \left( \begin{matrix}
         M_{11} & M_{12}\\
         M_{21} & M_{22}
    \end{matrix} \right)
    \begin{bmatrix}
        P_{xt}\\
        u_{xt}
    \end{bmatrix}
\end{equation}
where $M_{11}=M_{22}$ and $M_{11} M_{22}-M_{12} M_{21}=1$. In these equations, $P_{x0}$, $u_{x0}$, $P_{xt}$, and $u_{xt}$ are the overall pressure and velocity field at $x=0$ and $x=t$, respectively. The overall pressure and velocity field at $x=0$ and $x=t$ can be defined as:
\begin{equation}\label{eq4}
    P_{x0}=\frac{1}{S_2}  [S_1 \overline{P_1}(x=0)+S_3 \overline{P_2}(x=0)]
\end{equation}
\begin{equation}\label{eq5}
    P_{xt}=\frac{1}{S_2}  [S_1 \overline{P_1}(x=t)+S_3 \overline{P_2}(x=t)]
\end{equation}
\begin{equation}\label{eq6}
    u_{x0}=\frac{1}{S_2}  [\overline{U_1}(x=0)+ \overline{U_2}(x=0)]
\end{equation}
\begin{equation}\label{eq7}
    u_{xt}=\frac{1}{S_2}  [ \overline{U_1}(x=t)+ \overline{U_2}(x=t)]
\end{equation}
where $S_1=\pi r_1^2$, $S_2=\pi r_2^2$, and $S_3=\pi (r_2^2-r_1^2)$. In these equations, $\overline{P_1}(x=0)$, $\overline{P_2}(x=0)$, $\overline{P_1}(x=t)$), $\overline{P_2}(x=t)$, $\overline{U_1}(x=0)$, $\overline{U_2}(x=0)$, $\overline{U_1}(x=t)$, and $\overline{U_2}(x=t)$ are the averaged pressures and volume velocities at part 1 and 2 of region $B$ at $x=0$ and $x=t$, respectively. The mathematical definition of these averaged pressures and volume velocities are written in Supplementary Note A. 

It is shown that the averaged pressures $\overline{P_1}(x=0)$, $\overline{P_2}(x=0)$, $\overline{P_1}(x=t)$), and $\overline{P_2}(x=t)$ and volume velocities $\overline{U_1}(x=0)$, $\overline{U_2}(x=0)$, $\overline{U_1}(x=t)$ and $\overline{U_2}(x=t)$ can be calculated as (the detailed derivation process is given in Supplementary Note A):  
\begin{equation}\label{eq8}
    W=Q^{-1} Y 
\end{equation}
where $Y$, $Q$ and $W$ are defined as:
\begin{equation}\label{eq9}
    Y=\begin{bmatrix}
0 & 0 & 2 & 2 & 0 & 0 & 0 & 0
\end{bmatrix}^{T}
\end{equation}
\begin{widetext}
\begin{equation}\label{eq10}
    Q=\begin{pmatrix}
-\frac{S_1}{S_2} & -\frac{S_3}{S_2} & -\frac{M_{11}S_1}{S_2} & -\frac{M_{11}S_3}{S_2} & 0 & 0 & \frac{M_{12}}{S_2} & \frac{M_{12}}{S_2} \\
0 & 0 & \frac{M_{21}S_1}{S_2} & \frac{M_{21}S_3}{S_2} & -\frac{1}{S_2} & -\frac{1}{S_2} & \frac{M_{22}}{S_2} & \frac{M_{22}}{S_2} \\
1 & 0 & 0 & 0 & -A & -B & 0 & 0 \\
0 & 1 & 0 & 0 & -C & -D & 0 & 0 \\
0 & 0 & 1 & 0 & 0 & 0 & -E & -F \\
0 & 0 & 0 & 1 & 0 & 0 & -G & -H \\
0 & 1 & 0 & -cos(k_0 n_2 t) & 0 & 0 & 0 & iz_2sin(k_0 n_2 t) \\
0 & 0 & 0 & \frac{i}{z_2}sin(k_0 n_2 t) & 0 & 1 & 0 & cos(k_0 n_2 t)
\end{pmatrix}
\end{equation}
\begin{equation}\label{eq11}
    W=\begin{bmatrix}
        \overline{P_1}(x=0) & \overline{P_2}(x=0) & \overline{P_1}(x=t) & \overline{P_2}(x=t) & \overline{U_1}(x=0) & \overline{U_2}(x=0) & \overline{U_1}(x=t) & \overline{U_2}(x=t)
    \end{bmatrix}^{T}
\end{equation}
\end{widetext}
where $A, B, C, D, E, F, G$ and $H$ are as:
\begin{equation}\label{eq12}
    A=\frac{4i\rho_0 \omega}{r_1^4}\int_{0}^{r_1}\int_{0}^{r_1}G_1(r,0;r_0,0)r_0 dr_0 r dr
\end{equation}
\begin{equation}\label{eq13}
    B=\frac{4i\rho_0 \omega}{r_1^2\left( r_2^2-r_1^2 \right)}\int_{0}^{r_1}\int_{r_1}^{r_2}G_1(r,0;r_0,0)r_0 dr_0 r dr
\end{equation}
\begin{equation}\label{eq14}
    C=\frac{4i\rho_0 \omega}{r_1^2\left( r_2^2-r_1^2 \right)}\int_{r_1}^{r_2}\int_{0}^{r_1}G_1(r,0;r_0,0)r_0 dr_0 r dr
\end{equation}
\begin{equation}\label{eq15}
    D=\frac{4i\rho_0 \omega}{\left( r_2^2-r_1^2 \right)^2}\int_{r_1}^{r_2}\int_{r_1}^{r_2}G_1(r,0;r_0,0)r_0 dr_0 r dr
\end{equation}
\begin{equation}\label{eq16}
    E=-\frac{4i\rho_0 \omega}{r_1^4}\int_{0}^{r_1}\int_{0}^{r_1}G_2(r,t;r_0,t)r_0 dr_0 r dr
\end{equation}
\begin{equation}\label{eq17}
    F=-\frac{4i\rho_0 \omega}{r_1^2\left( r_2^2-r_1^2 \right)}\int_{0}^{r_1}\int_{r_1}^{r_2}G_2(r,t;r_0,t)r_0 dr_0 r dr
\end{equation}
\begin{equation}\label{eq18}
    G=-\frac{4i\rho_0 \omega}{r_1^2\left( r_2^2-r_1^2 \right)}\int_{r_1}^{r_2}\int_{0}^{r_1}G_2(r,t;r_0,t)r_0 dr_0 r dr
\end{equation}
\begin{equation}\label{eq19}
    H=-\frac{4i\rho_0 \omega}{\left( r_2^2-r_1^2 \right)^2}\int_{r_1}^{r_2}\int_{r_1}^{r_2}G_2(r,t;r_0,t)r_0 dr_0 r dr
\end{equation}

In Eqs.~\eqref{eq12}-\eqref{eq19}, $G_1(r,0;r_0,0)$ and $G_2(r,t;r_0,t)$ are Green's function at the interfaces and can be simplified as \cite{feng2013acoustic, li2014three}:
\begin{equation}\label{eq20}
    G_1(r,0;r_0,0)=G_2(r,t;r_0,t)=\sum_{n=0}^{n=\infty}\frac{\varphi_n(r_0)\varphi_n(r)}{-i\pi r_2^2 \sqrt{k_0^2-k_n^2}}
\end{equation}
where $\phi_n (r)$ represents the duct's eigenmode as:
\begin{equation}\label{eq21}
    \varphi_n (r)=\frac{J_0 (k_0 r)}{J_0 (k_n r_2) }
\end{equation}
when $J_0$ is the zeros order Bessel function and $k_n$ denotes the wavenumber, which is the solution of $J' (k_n r_2 )=0$.

Eventually, the effective refractive index ($n_1$) and effective impedance ($z_1$) of part 1 can be obtained from the calculated averaged pressures and volume velocities as (the detailed derivation process is given in Supplementary Note A):
\begin{equation}\label{eq22}
    z_1=\sqrt{\frac{\overline{P_1}(x=0)^2-\overline{P_1}(x=t)^2}{\overline{U_1}(x=0)^2-\overline{U_1}(x=t)^2}}
\end{equation}
\begin{equation}\label{eq23}
    n_1=\frac{\pm cos^{-1} \left( \frac{\overline{P_1}(x=0)\overline{U_1}(x=0)+\overline{P_1}(x=t)\overline{U_1}(x=t)}{\overline{P_1}(x=0)\overline{U_1}(x=t)+\overline{P_1}(x=t)\overline{U_1}(x=0)}\right)+2 \pi m}{k_0 t}
\end{equation}
where $m$ is the branch number of $cos^{-1}$ function

\section{Verification and Results} \label{sec3}
In this section, the proposed method in section~\ref{sec2} is evaluated with data from numerical simulation. To this end, acoustic wave propagation inside the duct and metamaterial is simulated by the Finite Element Method (FEM) in COMSOL Multiphysics. In this simulation, the thermoviscous effect is ignored, and a perfectly match layer assumption is taken to avoid reflecting from the sides of the duct. The metamaterial (part 1 of region $B$) is defined in the simulation with a known effective refractive index and effective impedance of $n_1$ and $z_1$. The complex transmission ($T$) and reflection ($R$) coefficients are measured utilizing the two-tube three-microphone method \cite{ho2005measurements}. Then, the measured complex transmission ($T$) and reflection ($R$) coefficients are imported to the proposed method in the previous section and $n_1$ and $z_1$ are calculated over frequency and compared with the defined $n_1$ and $z_1$ in the simulation. 

\begin{figure}[!t]
    \centering
    \includegraphics[width=3.4in,scale=1]{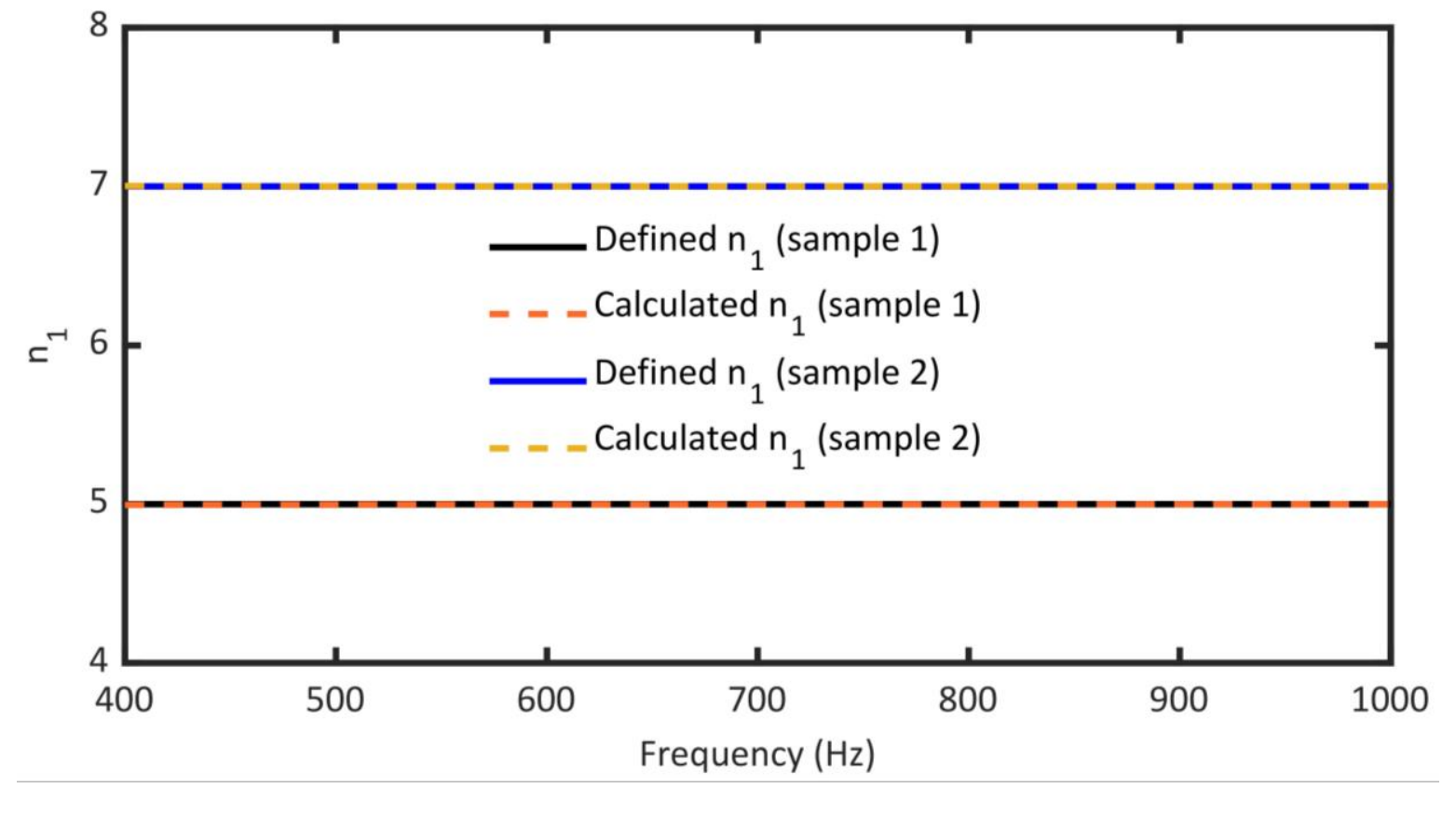}
    \caption{Extracted effective refractive indexes for sample 1 and 2.}
    \label{fig2}
\end{figure}

\begin{figure}[!t]
    \centering
    \includegraphics[width=3.4in,scale=1]{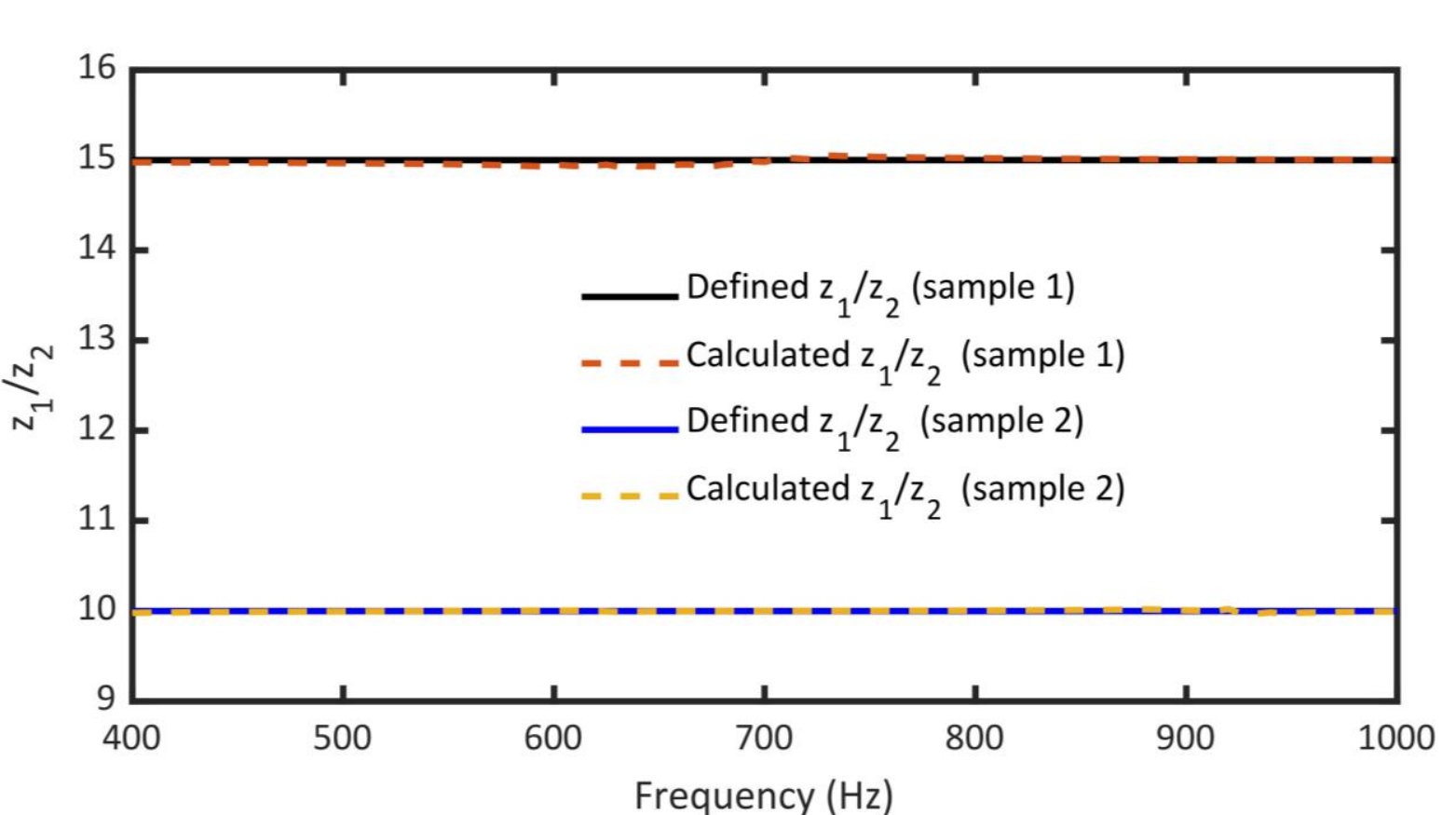}
    \caption{Extracted effective impedance ratio for sample 1 and 2.}
    \label{fig3}
\end{figure}

Herein, two samples with different dimensions and properties are evaluated. In sample 1, $r_1=40$~mm, $t=5.2$~mm, $n_1=5$, and $z_1/z_2=15$. In sample 2, $r_1=51$~mm, $t=8$~mm, $n_1=7$, and $z_1/z_2=10$. In these samples $r_2=70$~mm, $n_2=1$ and $z_2=\rho_0 c_0/\pi(r_2^2-r_1^2)$. The calculated effective refractive index ($n_1$) for the two samples are shown in Fig.~\ref{fig2}. As can be seen in this figure, the calculated $n_1$ is close to the defined $n_1$in the simulation. Also, the calculated effective impedance $z_1$ over $z_2$ ratio is plotted in Fig.~\ref{fig3}. As shown in this figure, the calculated effective impedance ratio ($z_1/z_2$) is matched with the defined effective impedance ratio with an error of below 1\%. The accuracy of calculated $n_1$ and $z_1$ indicates the applicability of the proposed method in extracting the effective properties of metamaterials when the impedance tube and metamaterial are not the same size.

\section{Conclusion} \label{sec4}
In this paper, we proposed an analytical methodology to extract the effective acoustic properties of metamaterials located inside an impedance tube when the metamaterial and duct do not have the same size, and there is an air gap. To this end, a bilayer metamaterial (part B) was assumed inside a duct to include the air gap with known effective acoustic properties. TMM is employed to calculate the averaged acoustic pressures and velocities at parts 1 and 2 of region B at $x=0$ and $x=t$ from the measured complex transmission and reflection coefficients. The proposed methodology showed that the effective properties of metamaterial could be obtained by solving a set of eight linear equations. FEM is used to simulate acoustic wave propagation inside the duct and metamaterial and measure the complex transmission and reflection coefficients. Finally, the effective properties of metamaterial were obtained and compared with defined values in different frequencies. The results demonstrated the proposed method's strong ability to provide an analytical solution to extract the effective properties of metamaterials with an accuracy of 99\% when the metamaterial and impedance tube possess different sizes. 

\section*{supplementary material}
See supplementary Note A for the complete derivation of mathematical model.

\section*{Acknowledgments}
The authors declare that they have no known competing for financial interests or personal relationships that could have appeared to inﬂuence the work reported in this paper.

\section*{Data availability}
The data that support the plots within this paper and other findings of this study are available from the corresponding author upon request.

\bibliographystyle{apsrev4-1} 
\bibliography{ref.bib} 

\end{document}